\documentclass[a4paper,11pt]{article}

\usepackage{pos}
\usepackage{graphicx}
\graphicspath{{pictures/}} 
\usepackage{tabularx}
\usepackage{floatrow}
\usepackage{subfig}
\usepackage{dsfont} 
\usepackage{bookmark}
\usepackage{braket}
\usepackage[outdir=pictures/]{epstopdf}

\title{Investigating \boldmath{$N \to N\pi$} axial matrix elements}

\author*{Lorenzo Barca}
\author{Gunnar S. Bali}
\author{Sara Collins}

\affiliation{Institute for Theoretical Physics, University of Regensburg\\
 93040,  Regensburg, Germany}

\emailAdd{lorenzo1.barca@ur.de, gunnar.bali@ur.de, sara.collins@ur.de}

\abstract{
Excited state contamination is one of the most challenging sources of systematics to tackle in the determination of nucleon matrix elements and form factors.
The signal-to-noise problem prevents one from considering large source-sink time separations
for the three-point functions to ensure ground state dominance.
Instead, relevant analyses consider multi-state fits.
Excited state contributions are particularly significant in the axial channel.
In this work, we confront the problem directly.  
Since the major source of contamination is understood to be related to pion production, 
we consider three-point correlation functions with a nucleon operator at the source 
and a nucleon-pion interpolating operator at the sink, which allows studies of $N \to N\pi$ matrix elements.
We discuss the construction of these three-point correlation functions and 
we solve the generalized eigenvalue problem (GEVP) using different sets of nucleon and nucleon-pion interpolators.
The analysis is performed on the CLS ensemble A653 with $m_\pi \approx 420$ MeV.
Results were generated with valence quark masses corresponding to 
$m_\pi \approx 1750$ MeV and $m_\pi \approx 420$ MeV.
}

\FullConference{%
 The 38th International Symposium on Lattice Field Theory, LATTICE2021
  26th-30th July, 2021
  Zoom/Gather@Massachusetts Institute of Technology
}

\begin{document}
\maketitle

\section*{Introduction}
The nucleon axial matrix elements $\bra{N'} {\cal{A}} \ket{N}$, 
as well as the nucleon vector matrix elements $\bra{N'} {\cal{V}} \ket{N}$, 
enter in the cross-sections of weak interaction processes, 
which are being studied in modern high precision experiments.
$N'$ could either be a nucleon or a nucleon-pion final state.
Both are produced, e.g., in the terrestrial long-baseline neutrino oscillation experiments.
In particular,  the Lorentz decomposition of the matrix elements into form factors is required.
On the lattice, it is possible to determine directly the nucleon to nucleon matrix elements
by computing suitable ratios of $N\to N$ three-point functions and $N\to N$ two-point functions.
However, it has been found that the extraction of these quantities is contaminated by excited states, in particular $N\pi$.
$\chi$PT analyses have shown that the closer we go to physical pion masses, 
the more the $N\pi$ states cannot be neglected, especially for the pseudoscalar form factors $G_P$ and $G_{\tilde{P}}$ (see \citep{Baer_2019_axial, Baer_2019_pseudoscalar}).
Recently, our group estimated the tree-level $\chi$PT $N\pi$ contribution to the ratio in the axial and pseudoscalar channels
and was able to extract form factors that satisfy the PCAC relation (see \cite{Bali_2020}).
In this work, we compute directly $N\to N\pi$ three-point functions 
and in the final analysis we aim to extract $N\to N\pi$ matrix elements,
with the aim of validating $\chi$PT expectations, but also to explore $\pi$ production.
\\
$~ \qquad$This paper is organised as follows.
In section 1, we discuss nucleon two-point functions and we present results of the GEVP using a basis of local and smeared interpolators.
In section 2, we discuss standard nucleon three-point functions and we present results of the nucleon axial matrix element
in the forward limit., i.e., $g_A$.
In section 3, we construct the nucleon-pion interpolators at the sink (see \cite{Prelovsek_2017}) 
in order to compute $N\to N\pi$ three-point functions 
and we discuss how we obtain $N\pi$ two-point functions from them.
In the final section, we present results of the GEVP applied to a $3\times3$ matrix 
of two-point correlation functions using a basis of local, 
smeared nucleon interpolators and nucleon-pion interpolators. 
Previous works using $N\pi$ operators include studies of $N\pi$ scattering (\citep{Kiratidis:2015vpa, Leskovec:2018lxb}).
The analysis is performed on the CLS ensemble $A653$ with $L_s=24~a$,  $L_t = 48~a$, $\beta=3.34$, $m_\pi \approx 420$ MeV, lattice spacing $a\approx 0.1$ fm and (anti) periodic boundary conditions in time for the (fermion) gauge fields.

\section{Nucleon two-point functions}
\begin{figure}
 \includegraphics[width=7.4cm]{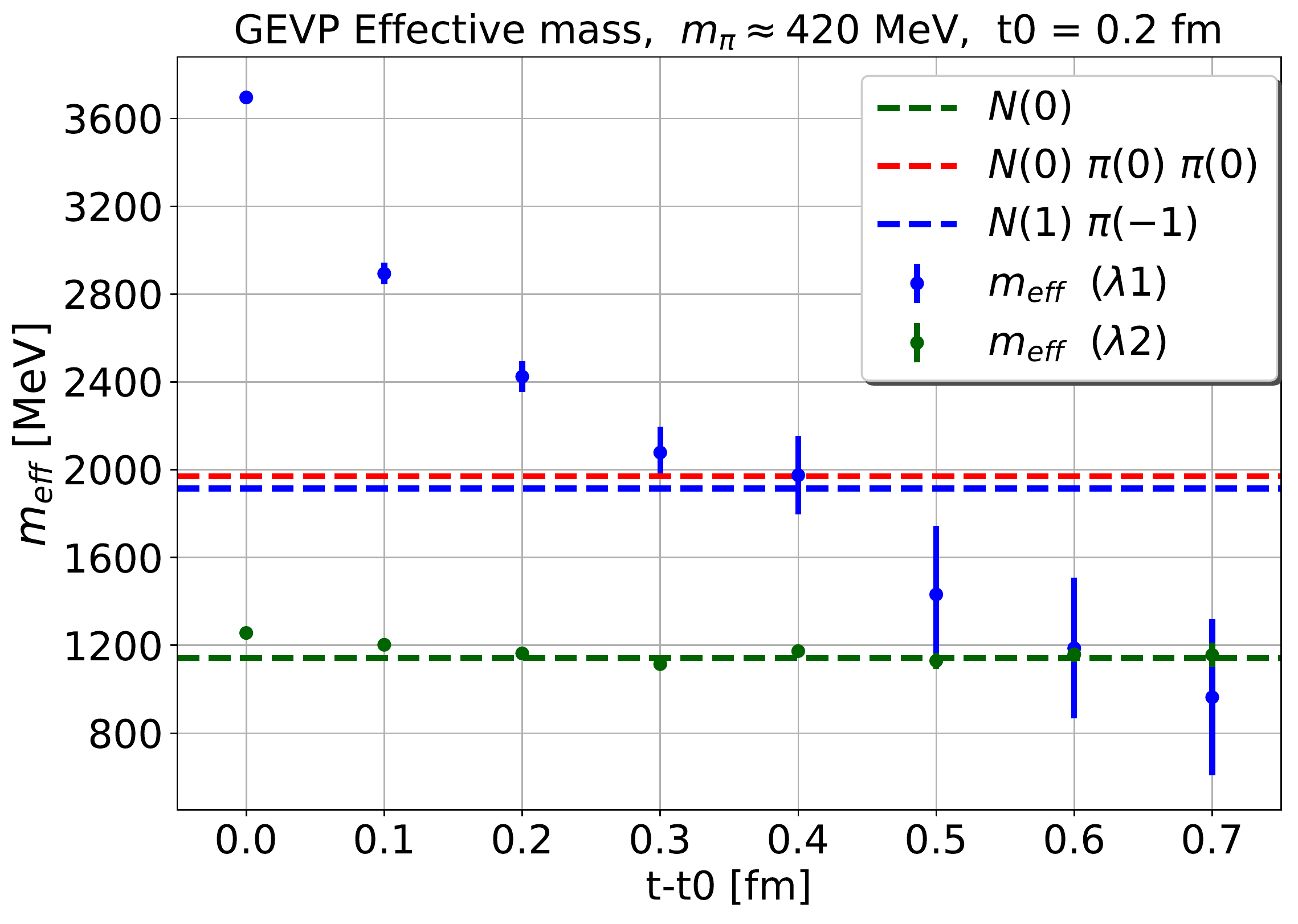}
 \hspace{0.1cm}
 \includegraphics[width=7.4cm]{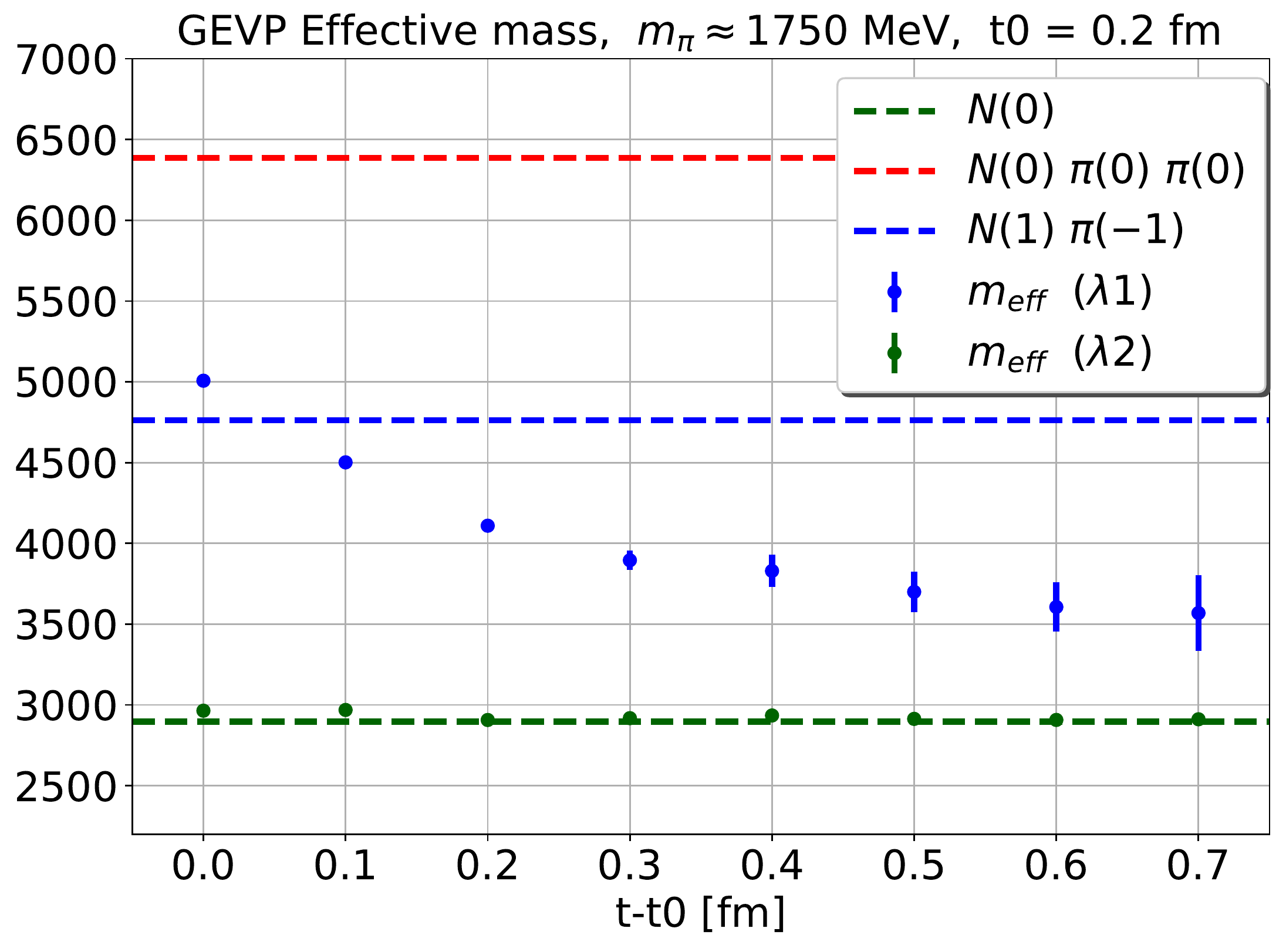}
   \caption{ \\
   Effective masses in the rest-frame, extracted from the two eigenvalues of the GEVP at a reference time $t_0=0.2$ fm using the basis \{${\cal{O}}_N$,  $\Phi {\cal{O}}_N$\}.
   The blue and red dashed lines represent, respectively, 
   the non-interacting energy of a $N\pi$ in a P-wave and 
   $N\pi\pi$ in a S-wave.  }
    \label{figure:meff_nucleon2pt}
\end{figure}

We define the nucleon annihilation operator at the sink $x \equiv (\vec{x}, t)$:
\begin{align}\label{nucleon_interpolator1}
{\cal{O}}_N^\gamma(x)~ = ~\epsilon^{abc}~ \Bigl( u_a(x) ~ C\gamma_5~ d_b(x)\  \Bigr)~ q^\gamma_{c}(x),
\end{align}
where $q = u,  d$ depending on which isospin component of the nucleon we consider $N=p, n$.\\
The nucleon 2-point function is defined as 
\begin{equation}\label{2-point_function}
C^{\vec{p}}_{2pt}(t, t_0) ~ = ~P^+_{\bar{\gamma}\gamma} ~ \sum_{\vec{x}} ~ e^{-i \vec{p}\cdot (\vec{x}-\vec{x}_0) } ~\braket{~{\cal{O}}_N^\gamma(\vec{x}, t) ~ \bar{{\cal{O}}}_N^{\bar{\gamma}}(\vec{x_0}, t_0)~}
\end{equation}
where $\bar{{\cal{O}}}_N$ is the nucleon creation operator at a fixed source position $x_{0}\equiv(\vec{x}_{0}, t_{0})=(\vec{0},0)$, $P^+$ is the positive-parity projector and the Fourier transform in the spatial coordinates allows the interpolators to have a definite momentum $\vec{p} = \frac{2\pi}{L} \vec{n}$.
\\
\hfill \break
Employing the spectral decomposition by inserting a complete set of states $\sum_n \frac{1}{2E_n} \ket{n} \bra{n}$ in to Eq. \eqref{2-point_function}, we obtain
\begin{equation}\label{spectral_nucleon2pt}
C^{\vec{p}}_{2pt}(t)~=
~\sum_{\sigma}~P^+_{\bar{\gamma}\gamma}  ~ \frac{e^{-E_N^{\vec{p}} t}}{2 E_N^{\vec{p}}}
\bra{\Omega} {\cal{O}}_N^{\vec{p}, \gamma} \ket{N_\sigma^{\vec{p}}} \bra{N_\sigma^{\vec{p}}} \bar{{\cal{O}}}_N^{\bar{\gamma}} \ket{\Omega} + ...
\xrightarrow{\text{$t\gg 0$}}~
|Z_{\vec{p}}|^2 ~ \frac{E_N^{\vec{p}}+m_N}{m_N} e^{-E_N^{\vec{p}}t},
\end{equation}
where $ {\cal{O}}_N^{\vec{p}, \gamma}$ is the nucleon interpolator, projected onto momentum $\vec{p}$.
In principle, each state with the same quantum numbers as the nucleon contributes to the two-point function 
and this includes radially excited nucleons and multiparticle states.
The overlap factors $Z_{\vec{p}}$ are defined as $\bra{\Omega} {\cal{O}}_N \ket{N_\sigma^{\vec{p}}} = u_N(\vec{p}, \sigma) Z_{\vec{p}}$.
At large source-sink separation, we expect the nucleon ground state to be dominant, 
so that for the momentum $\vec{p}=\vec{0}$, we can extract the nucleon mass $m_N = E_N^{\vec{p}=\vec{0}}$.
\\
\hfill \break
As a first step, we construct a $2\times 2$ matrix of correlators ${\cal{M}}(t)$ using the basis \{${\cal{O}}_N$, $\Phi{\cal{O}}_N$\},
where $\Phi{\cal{O}}_N$ is a spatially extended nucleon interpolator obtained by applying a Wuppertal smearing operator
$\Phi$ (\citep{GUSKEN1990361}), with APE-smeared gauge links (\citep{FALCIONI1985624}), to smooth the quarks $q$.
The smearing is applied either only at the source or both at the source and at the sink.
We solved the GEVP
\begin{equation}\label{gevp_equation}
{\cal{M}}(t)~V(t,t_0) ~ = ~ {\cal{M}}(t_0) ~V(t,t_0) ~ \Lambda(t,t_0)
\end{equation}
and extract the effective masses of the eigenvalues at a reference time $t_0$.
The results are displayed in Fig. \ref{figure:meff_nucleon2pt}.
The pion and nucleon masses that are needed to draw the dashed lines are extracted at large times from the pion and nucleon two-point functions (see Section $4$ for the definition of the pion interpolators).
In the heavier quark mass case ($m_\pi\approx1750$ MeV) there is evidence of a state between the nucleon ground state and the non-interacting $N\pi$ in a P-wave.

\section{Nucleon three-point functions and matrix elements}
With the insertion of an iso-vector\footnote{In this notation, $\tau^k$ are the Pauli matrices acting in flavor space and $\psi$ is the isospin doublet of $u$ and $d$ quarks.} 
current ${\cal{J}}^k \equiv \bar{\psi} \Gamma \frac{\tau^k}{2} \psi $ 
at a space-time position $z \equiv (\vec{z}, \tau)$ between the nucleon interpolators, we compute the nucleon three-point functions
\begin{equation}\label{nucleon_3pt}
C^{{\cal{J}},  P}_{3pt}(\tau; t; \vec{p}';\vec{q})~=~ P_{\bar{\gamma}\gamma}~\sum_{\vec{x}, \vec{z}} ~ e^{-i \vec{p}' \cdot \vec{x}} ~ e^{i \vec{q}\cdot \vec{z}}~ 
 \braket{~{\cal{O}}_N^\gamma(\vec{x}, t) ~  {\cal{J}}^k(\vec{z}, \tau) ~ \bar{{\cal{O}}}_N^{\bar{\gamma}}(\vec{0}, 0)~}.
\end{equation}
The ground state contribution to the spectral decomposition reads
\begin{align}\label{spectral_nucleon3pt}
C^{{\cal{J}},  P}_{3pt}(\tau)~=P_{\bar{\gamma}\gamma}
 \sum_{\sigma'\sigma} 
 &
 ~\frac{e^{-(E^{\vec{p}}_N -E^{\vec{p}'}_{N'})\tau} }{4E^{\vec{p}'}_{N'}E^{\vec{p}}_N} e^{-E^{\vec{p}'}_{N'} t}
\bra{\Omega}{\cal{O}}_n^{\vec{p}',\gamma}\ket{N'^{\vec{p}'}_{\sigma'}}
\bra{N'^{\vec{p}'}_{\sigma'}} {\cal{J}}^k \ket{N^{\vec{p}}_{\sigma}}
\bra{N^{\vec{p}}_{\sigma}} \bar{{\cal{O}}}_n^{\bar{\gamma}} \ket{\Omega}
+ ...
\end{align}
The dependence of the nucleon matrix elements $\bra{N'^{\vec{p}'}_{\sigma'}} {\cal{J}}^k \ket{N^{\vec{p}}_{\sigma}}$ 
on the initial and final momentum is characterized by the form factors.
In the case of an axial current ${\cal{A}}_{\mu}^k$ with $\Gamma_\mu = \gamma_\mu \gamma_5$ and a pseudoscalar current $ {\cal{P}}^k$ with $\Gamma = \gamma_5$,  the Lorentz decompositions read
\begin{align}
&
\label{axial_matrix_element}
\bra{N'^{\vec{p}'}_{\sigma'}}  {\cal{A}}_{\mu}^k \ket{N^{\vec{p}}_{\sigma}}~=~
\bar{u}_N(\vec{p}', \sigma') \Bigl[ \gamma_\mu \gamma_5 G_A(Q^2) - i\frac{Q_\mu}{2m_N} G_{\tilde{P}}(Q^2) \Bigr] \frac{\tau^k}{2}  u_N(\vec{p}, {\sigma}),
\\
&\bra{N'^{\vec{p}'}_{\sigma'}}  {\cal{P}}^k \ket{N^{\vec{p}}_{\sigma}}~=~
\bar{u}_N(\vec{p}', \sigma') ~ G_P(Q^2) \frac{\tau^k}{2} u_N(\vec{p}, {\sigma}),
\label{pseudoscalar_matrix_element}
\end{align}
where the functions $G_X(Q^2)$ with $X$ $\in$ $\{A, P, \tilde{P} \}$ are the axial, pseudoscalar and induced pseudoscalar form factors.
\begin{figure}[tp]
\floatbox[{\capbeside\thisfloatsetup{capbesideposition={right, center},capbesidewidth=3.5cm}}]{figure}[\FBwidth]
{\caption{
\\
Renormalized ratio $\tilde{R}_{g_A}$ of Eq. \eqref{ratio_a_forward_limit}
using different source-sink separations $t$ with point nucleon interpolators and 3 source-sink separations with smeared nucleon interpolators.} \label{fig:figure_ratio_ga_tsep}}
{\includegraphics[width=11cm]{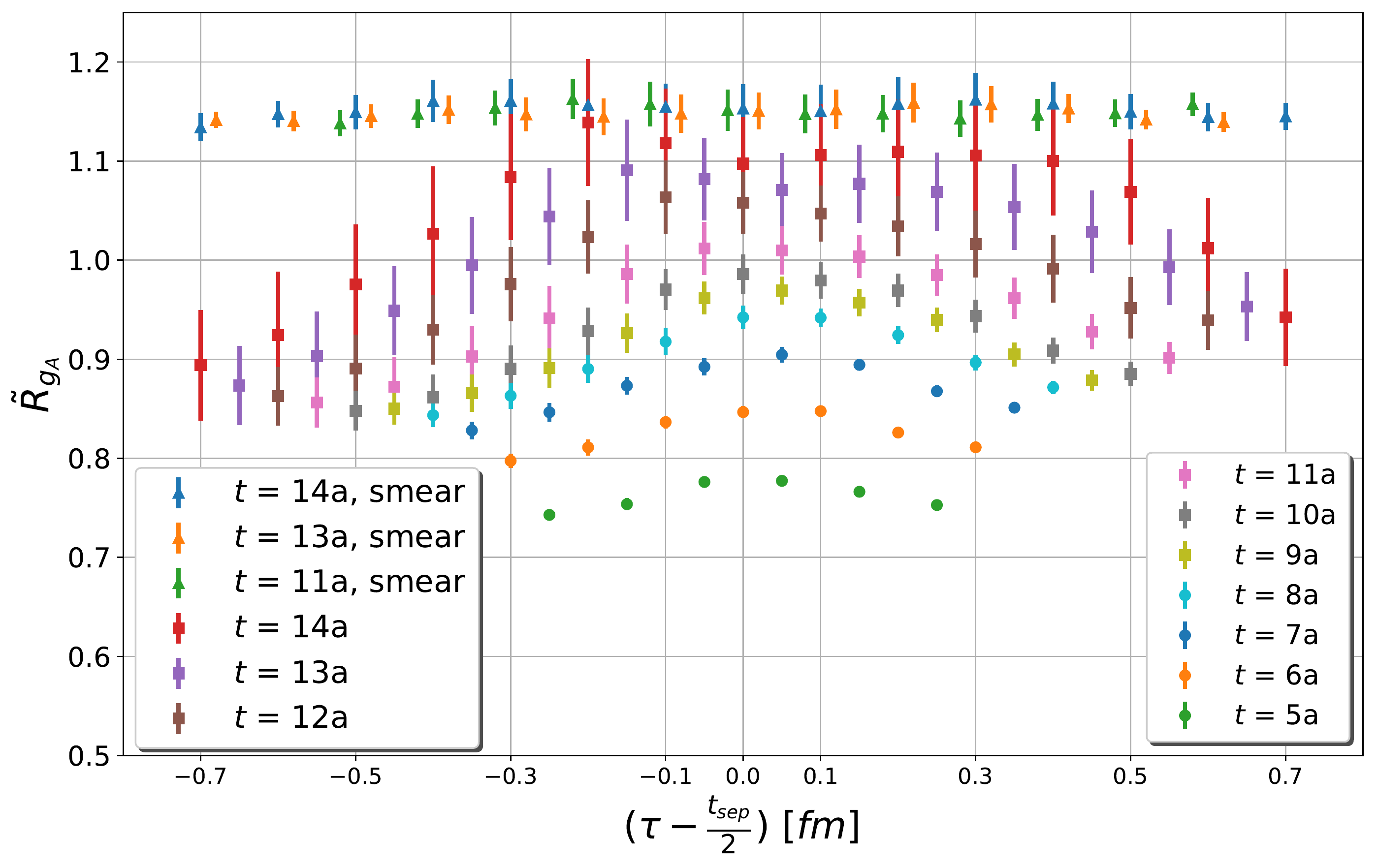}}
\end{figure}
It is possible to show through the spectral decomposition of the three-point function (see Eq. \eqref{spectral_nucleon3pt}) and of the two-point function (see Eq. \eqref{spectral_nucleon2pt}), 
that the (nucleon) ground state contribution to the ratio 
\begin{equation}\label{ratio_method}
R^{{\cal{J}}, P_i}_{\vec{p}', \vec{q}}(\tau;t) ~ = ~ \frac{C_{3pt}^{{\cal{J}}, P_i}(\tau; t; \vec{p}'; \vec{q}) }{C_{2pt}(t; \vec{p}' )} 
\sqrt{
\frac{C_{2pt}(\tau; \vec{p}') ~ C_{2pt}(t; \vec{p}' ) ~ C_{2pt}(t-\tau; \vec{p} )}
{C_{2pt}(\tau; \vec{p} ) ~C_{2pt}(t; \vec{p} )~ C_{2pt}(t-\tau; \vec{p} )}}
\end{equation}
is time-independent and that the overlap factors cancel.
In particular, in the forward-limit with $\vec{p}'= \vec{q} = \vec{0}$,  for the axial current we extract the unrenormalised axial charge $g_A$ of the nucleon:
\begin{equation}\label{ratio_a_forward_limit}
R_{g_A} (\tau) \equiv
R^{{\cal{A}}_i, P_i}_{\vec{0}, \vec{0}}(\tau;t) 
~\xrightarrow{t \gg \tau \gg 0} ~G_A(Q^2=0) \equiv g_A.
\end{equation}
In this limit, the ratio should not depend on $\tau$, however, due to excited states or multiparticle states contamination, 
there is a clear time-dependence,  especially when using point interpolators.
\\
We include in the analysis the renormalization factor 
$Z_A=0.7456 (10)^{\rm{stat}}(57)^{\rm{syst}}$, that was computed in \cite{Bali:2020lwx}
and we show in Fig. \ref{fig:figure_ratio_ga_tsep} the renormalized
ratio $\tilde{R}_{g_A}=Z_A R_{g_A}$,  
using the lighter quark mass ($m_\pi \approx 420$ MeV) and different source-sink separations $t$ without smearing 
and using smeared operators with $t=11 a, ~13a, ~14a$, where $a\approx 0.1$ fm.
The former ratios have a contribution from excited states which decreases with $t$, 
while in the latter cases the ground state appears to dominate within our statistical errors already for $t=11a$.
At $Q^2 \neq 0$,  the ground state contribution to Eq. \eqref{ratio_method} is proportional to the combination of nucleon form factors in Eq. \eqref{axial_matrix_element} and \eqref{pseudoscalar_matrix_element}.
A useful check that the form factors have been extracted correctly is provided by the PCAC relation
\begin{equation}
\frac{m_l^{PCAC}}{m_N} G_P(Q^2) ~ = ~ G_A(Q^2) - \frac{Q^2}{4m_N^2} G_{\tilde{P}}(Q^2) + {\cal{O}}(a^2),
\end{equation}
which should hold up to lattice artefacts. 
Many groups found the PCAC relation to be violated when performing a standard analysis,
where the excited state spectrum is determined from the two-point function alone.
However, it turns out (\citep{Baer_2019_axial, Baer_2019_pseudoscalar,  Bali_2020, gupta_axial_2020}) 
that the contribution of intermediate $N\pi$ state is enhanced in the three-point functions with respect to the two-point functions, 
which is expected by $\chi$PT (\citep{Baer_2019_axial, Baer_2019_pseudoscalar,  Bali_2020}).

\section{$N \to N\pi$ three-point functions}
\begin{figure}[]
 \includegraphics[width=6.5 cm]{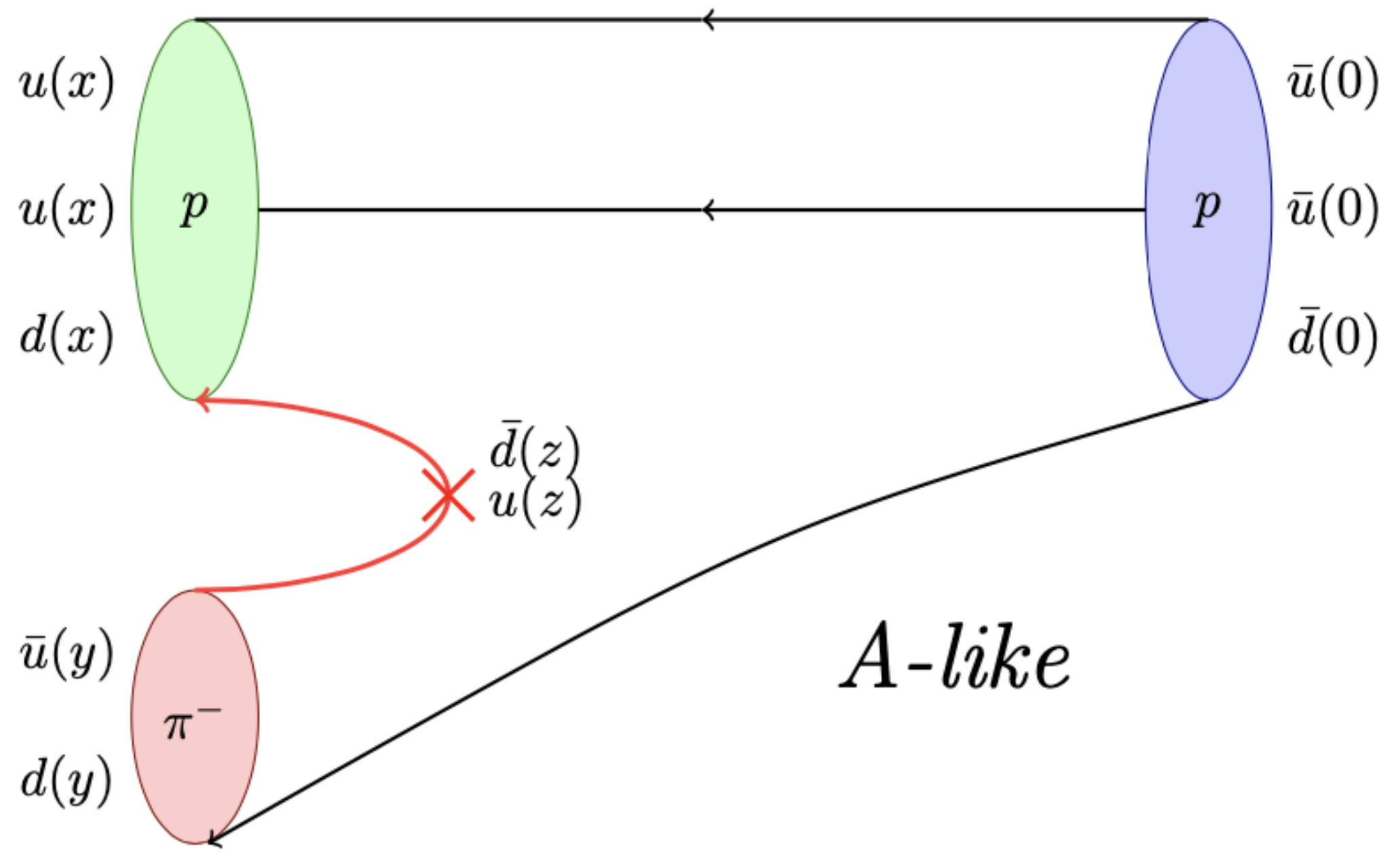}
 \hspace{1. cm}
 \includegraphics[width=6.5 cm]{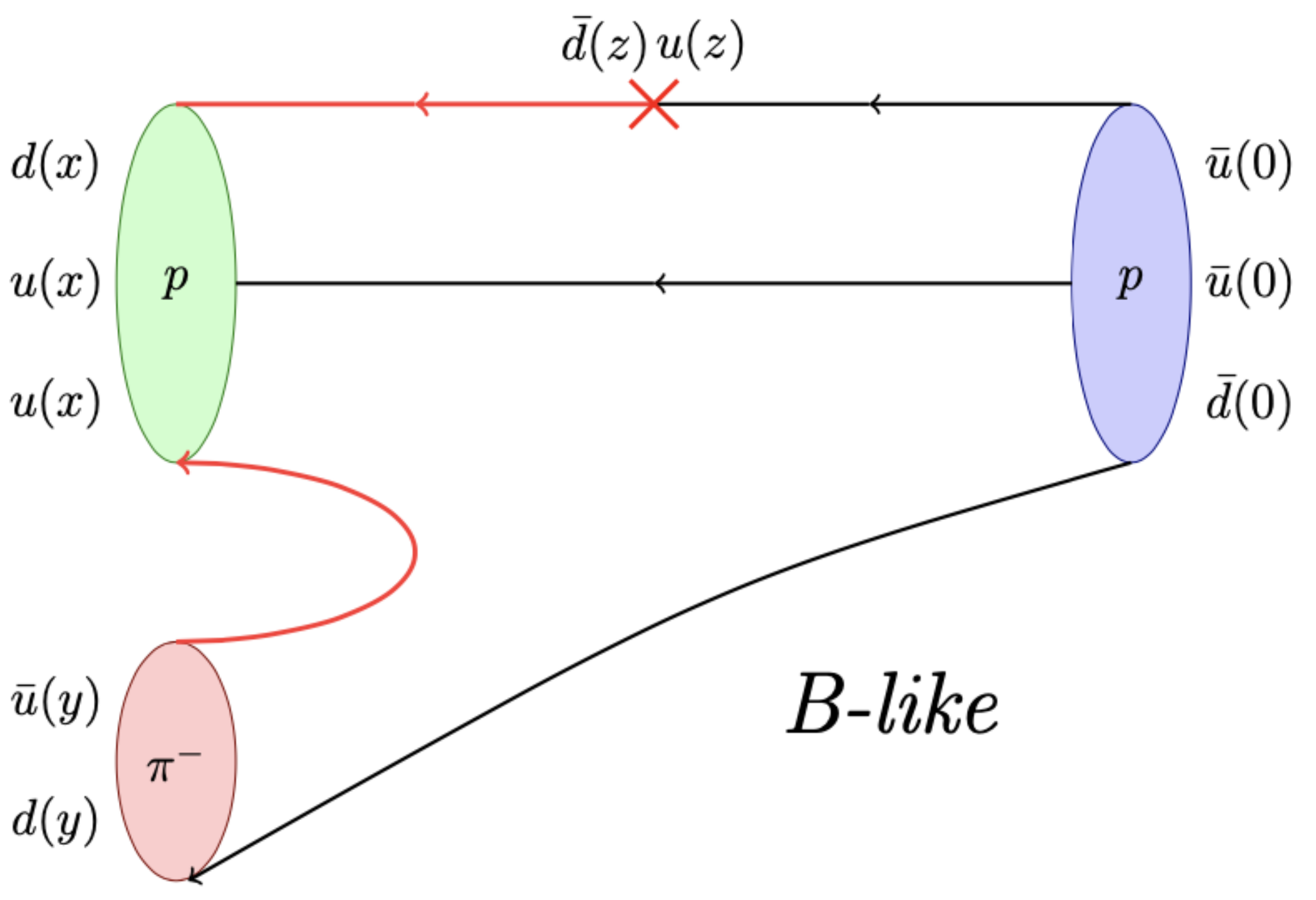} 
 \includegraphics[width=6.5 cm]{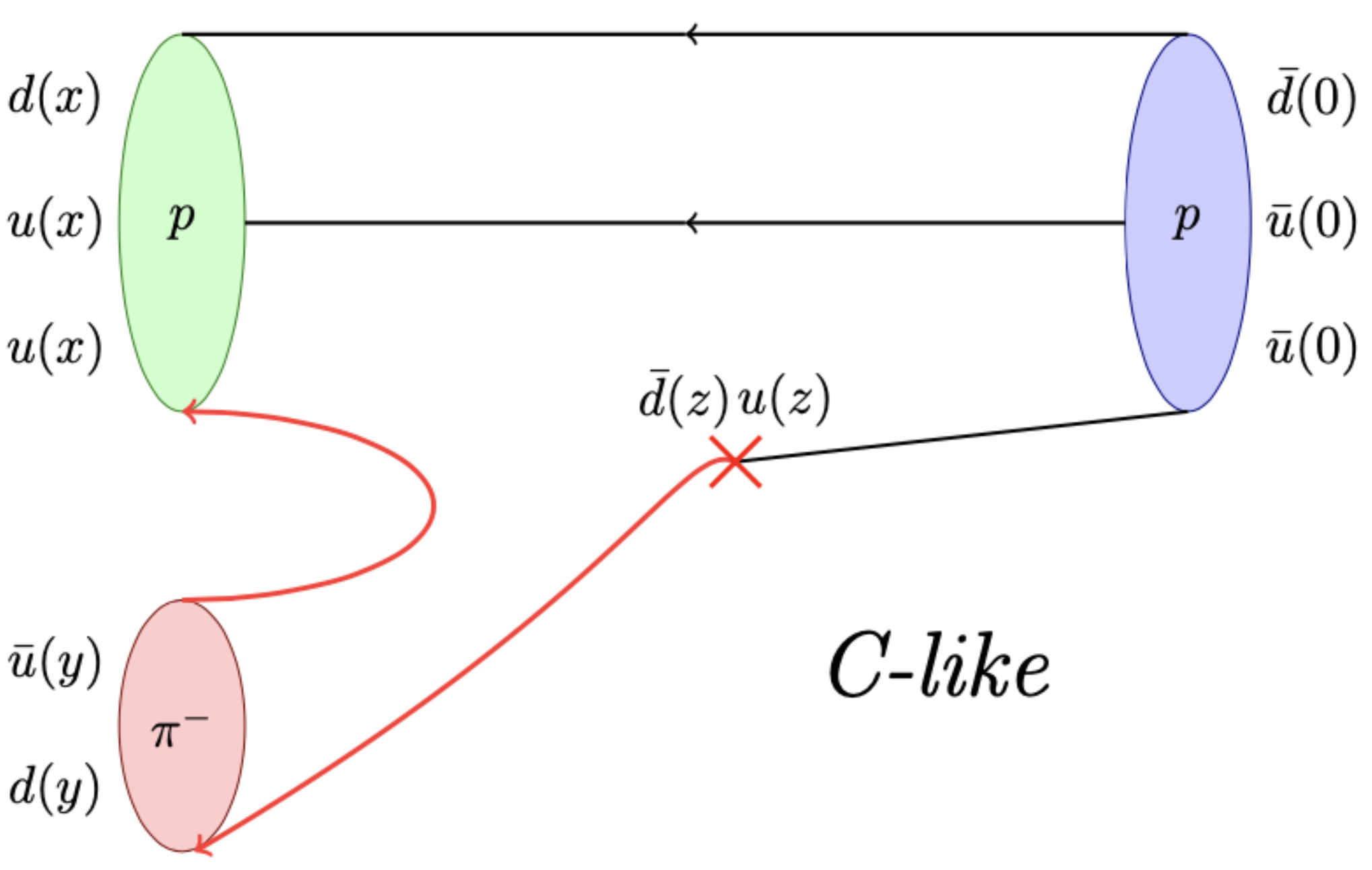}
 \hspace{1. cm}
 \includegraphics[width=6.5 cm]{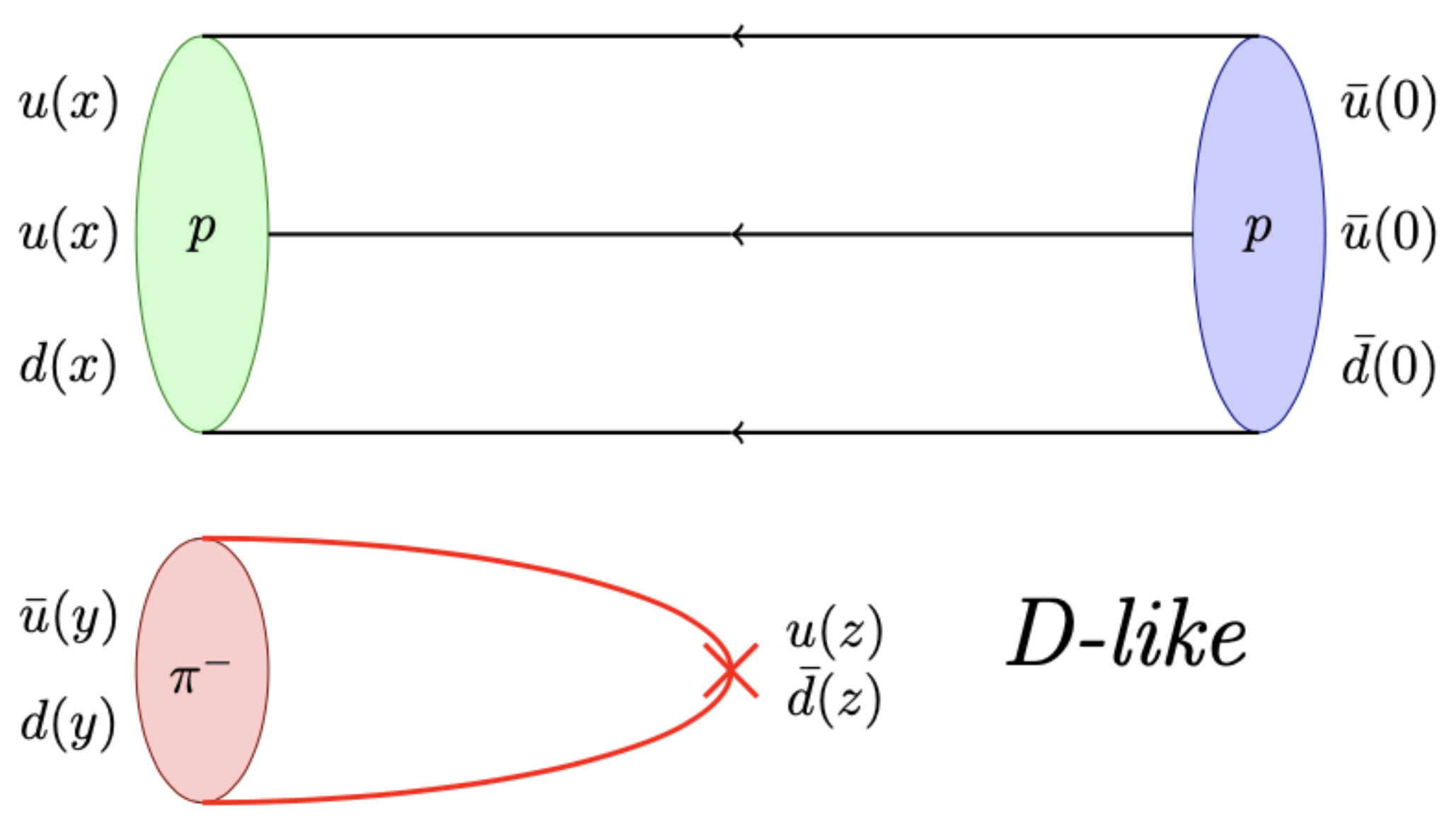} 
   \caption{ Sketches of the 4 different diagrams that emerge in the process $p \to p\pi^-$ via a ${{\cal{J}}^-}$ current.
   \\
   The total number of Wick contractions grows factorially with the number of quark-antiquark pairs. \\
   In this case there are in total $3$ u-quark pairs and $2$ d-quark pairs, which gives $3! 2! =12$ Wick contractions. 
   In particular, there are $2!$ permutations that are connected like the $A$,  
   $2! 2!$ like the $B$,  $2! 2!$ like the $C$ and $2!$ like the $D$ diagram.
   }
    \label{figure:diagrams_p2ppi}
\end{figure}

We compute for the first time the following three-point function:
\begin{equation}
C_{3pt}^{N \xrightarrow{{\cal{J}}} N\pi}(\tau, t)~=~
P_{\bar{\gamma}\gamma} \sum_{\vec{x}, \vec{y}, \vec{z}}
e^{-i \vec{p}'_N \cdot \vec{x}} ~e^{-i \vec{p}'_{\pi} \cdot \vec{y}} ~ e^{i \vec{q}\cdot \vec{z}}~ 
\braket{ ~ {\cal{O}}_{N}^{\gamma}(\vec{x},t)~{\cal{O}}_{\pi}(\vec{y},t)  ~ {\cal{J}}(\vec{z}, \tau) ~\bar{\cal{O}}_N^{\bar{\gamma}}(\vec{0}, 0) ~ },
\end{equation}
where at the sink timeslice $t$ we have the nucleon operator $ {\cal{O}}_{N}$ with momentum projection $\vec{p}'_N$ 
and the pion operator $ {\cal{O}}_{\pi}$ with momentum projection $\vec{p}'_\pi$.
The nucleon-pion interpolating operator must be projected onto the lattice irreducible representation $G_1$ and to the proper isospin $I=1/2$, $I_z = \pm 1/2$ depending on which isospin component of the nucleon we study.
For this study we focus on the neutron with $\ket{I=\frac{1}{2}, I_z=-\frac{1}{2}}$, 
corresponding to $P_{\bar{\gamma}\gamma}~\braket{~{\cal{O}}_n^\gamma(t)\ {\cal{J}}^{-}(\tau)\ \bar{{\cal{O}}}_p^{\bar{\gamma}}(0)\ }$,
where ${\cal{J}}^{-} = \frac{1}{2}({\cal{J}}^1-i{\cal{J}}^2)$.
Through the Wigner-Eckart theorem,  one can show that this correlation function is equivalent to 
the standard one $\braket{~{\cal{O}}_N^\gamma(t)\ {\cal{J}}^{3}(\tau)\ \bar{{\cal{O}}}_N^{\bar{\gamma}}(0)\ }$
of Eq. \eqref{nucleon_3pt}.
In particular, one can write a relation between scattering processes $S_{p \to n\pi^+}~=~\sqrt{2} S_{p \to p\pi^0}$
as a relation between three- or two-point functions,  recalling that $ {\cal{J}}^{\pi^0} = \sqrt{2}~ {\cal{J}}^3$.
The state $\ket{I=\frac{1}{2}, I_z=-\frac{1}{2}}$ is obtained from the isospin addition $I_1=1 ~ \oplus I_2=\frac{1}{2}$, which is
\begin{equation}\label{isospin_addition}
\ket{I=\frac{1}{2}, I_z=-\frac{1}{2}} ~=~ 
\frac{1}{\sqrt{3}} \ket{n} \ket{\pi^0}  
- \sqrt{ \frac{2}{3} } \ket{p} \ket{\pi^-},
\end{equation}
where we have identified the proton and the neutron as components of the isodoublet $I=1/2$ 
and the pions as components of the isotriplet $I=1$.  In particular,  the pion operators are
${\cal{O}}_{\pi^+}=-\bar{d} \gamma_5 u$,  ${\cal{O}}_{\pi^-}=\bar{u} \gamma_5 d$
and 
${\cal{O}}_{\pi^0}= \frac{1}{\sqrt{2}} \bigl(\bar{u} \gamma_5 u - \bar{d} \gamma_5 d\bigr)$.
\\
Considering Eq. \eqref{isospin_addition},  we also need to compute the following three-point functions: 
\\
$\braket{~{\cal{O}}_{p\pi^-}(t)~ {\cal{J}}^{-}(\tau)\ \bar{{\cal{O}}}_p(0)\ }$, 
$ \braket{~{\cal{O}}_{n\pi^0}(t)~ {\cal{J}}^{-}(\tau)\ \bar{{\cal{O}}}_p(0)\ }$
and
$\braket{~{\cal{O}}_{n\pi^0}(t)~ {\cal{J}}^{-}(\tau)\ \bar{{\cal{O}}}_p(0)\ }$.
\\
A useful check is provided by another Wigner-Eckart relation which must be fulfilled at the level of the correlators and matrix elements:
\begin{equation}\label{npi_wigner_eckart}
S_{n \pi^0 \to n\pi^0}~=~ S_{p \pi^- \to p\pi^-} + \frac{1}{\sqrt{2}}~ S_{p \pi^- \leftrightarrow n\pi^0}.
\end{equation}
In the case of $p \xrightarrow{\text{${\cal{J}}^-$}} p \pi^-$, there are $12$ Wick contractions to perform, 
that can be grouped in four different diagrams,  which we denote $A$, $B$, $C$ and $D$ (see Fig. \ref{figure:diagrams_p2ppi}).

\begin{figure}[tp]
\floatbox[{\capbeside\thisfloatsetup{capbesideposition={right, center},capbesidewidth=4.1 cm}}]{figure}[1.\FBwidth]
{\caption{
\\
   Sum of the diagrams A, B and C 
   of the three-point function $ P_i \braket{~{\cal{O}}_{n\pi^0}(t)~ {\cal{A}}^{-}_1(\tau)\ \bar{{\cal{O}}}_p(0)\ }$ 
   with $m_\pi \approx1750$ MeV at $t=1.7 ~\rm{fm}$, 
   using different polarizations $P_i$ and some of the momentum combinations that are needed to project $O_{n\pi^0}$ to the rest-frame.
} \label{figure:plots_NtoNpi_3pt}}
{\includegraphics[width=10cm]{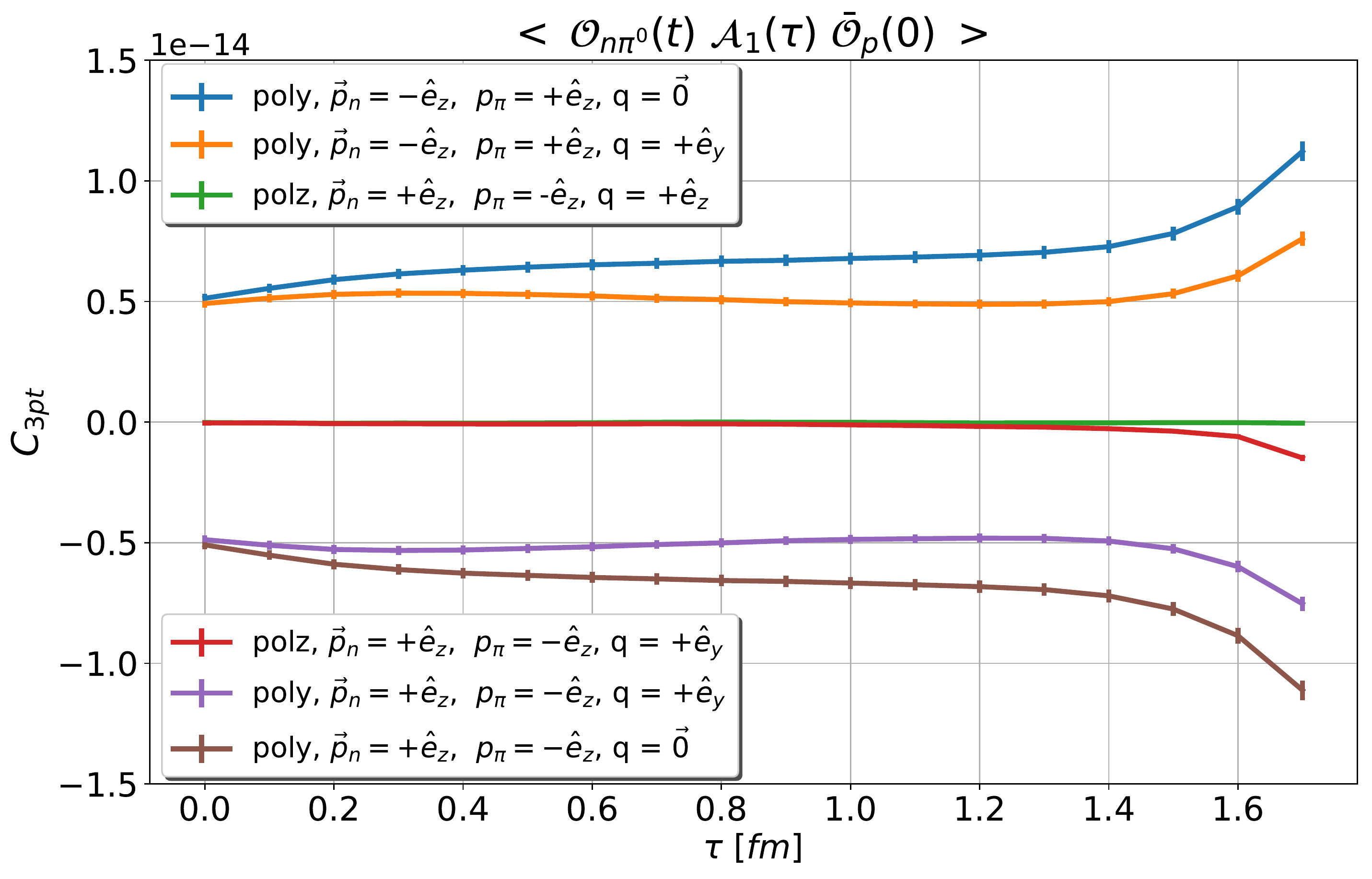}}
\end{figure}

Similar Wick contractions are computed for the other processes 
($p \xrightarrow{\text{${\cal{J}}^-$}} n \pi^0$ and $n \xrightarrow{\text{${\cal{J}}^3$}} n \pi^0$)
and we cross-check our expression against the Mathematica library QCT \cite{Djukanovic_2020} 
and our numerical results against the Wigner-Eckart relation.
The diagrams $A$, $B$ and $C$ are computed with the sequential source method \citep{Martinelli:1988rr}.
In particular, the red lines in Fig. \ref{figure:diagrams_p2ppi} represent all-to-all propagators, that we compute sequentially,
while the black lines are point-to-all propagators and the red crosses represent the insertion current ${\cal{J}}^-$.
The nucleon part of the diagram $D$ is computed with the point-to-all method, 
while the meson part with the One-End trick \citep{Foster_1999}.
In Fig. \ref{figure:plots_NtoNpi_3pt},  for the heavy quark mass, we plot a specific channel of 
$ P_i \braket{~{\cal{O}}_{n\pi^0}(t)~ {\cal{J}}^{-}(\tau)\ \bar{{\cal{O}}}_p(0)\ }$, 
where ${\cal{J}} = {\cal{A}}_1$ and the total sink momentum is zero.
The ${\cal{O}}_{N\pi}$ interpolator at the sink must be projected onto the lattice irreducible representation $G_1$. 
In the rest-frame $\vec{p}=\vec{p}'_N + \vec{p}'_\pi = \vec{0}$.
For $S_z=+1/2$, considering particles with a unit lattice momentum,  the projection onto $G_1$ reads:
\begin{align}\label{projection_g1_restframe}
\nonumber
{\cal{O}}^{N\pi, G_1}_{\vec{p}=\vec{0}, m_s=1/2}~=~
&
{\cal{O}}_N^{-1/2}(-\hat{e}_x) {\cal{O}}_\pi (\hat{e}_x)
~-
{\cal{O}}_N^{-1/2}(\hat{e}_x) {\cal{O}}_\pi (-\hat{e}_x)
~-i
{\cal{O}}_N^{-1/2}(-\hat{e}_y) {\cal{O}}_\pi (\hat{e}_y)
~+
\\
&
+i
{\cal{O}}_N^{-1/2}(\hat{e}_y) {\cal{O}}_\pi (-\hat{e}_y)
+
{\cal{O}}_N^{1/2}(-\hat{e}_z) {\cal{O}}_\pi (\hat{e}_z)
-
{\cal{O}}_N^{1/2}(\hat{e}_z) {\cal{O}}_\pi (-\hat{e}_z),
\end{align}
where $\hat{e}_j$ are the unit lattice momentum vectors (e.g., $\hat{e}_x = \frac{2\pi}{L} (1,0,0)$).
In the next section, we discuss how to compute the $N\pi$ two-point functions using the $N\to N\pi$ three-point functions evaluated at different source-sink separations.

\section{$N \pi$ two-point functions}
\begin{figure}[]
\begin{minipage}[]{1.\textwidth}
 \includegraphics[width=7.4cm]{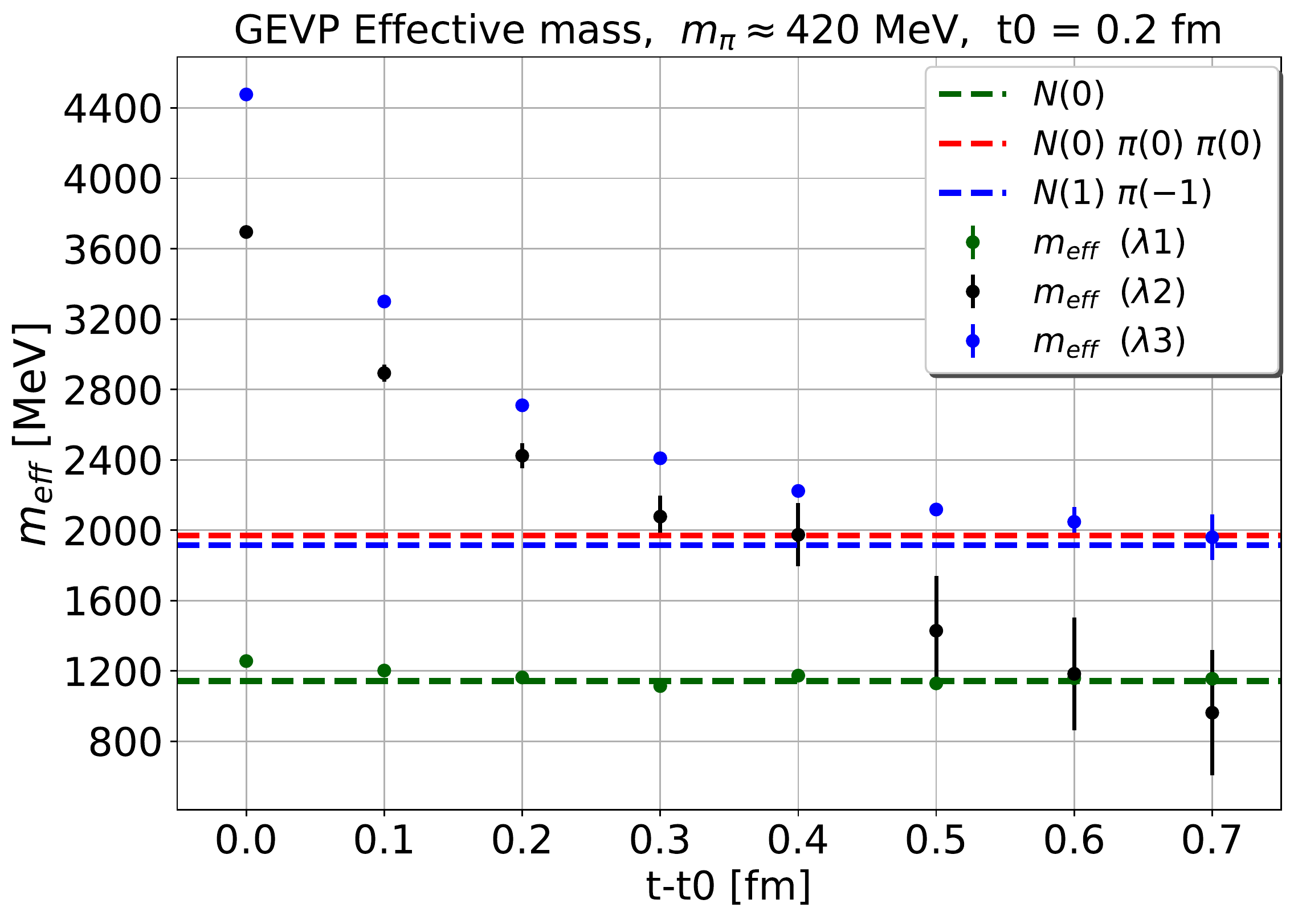}
 \hspace{0.1 cm}
 \includegraphics[width=7.4cm]{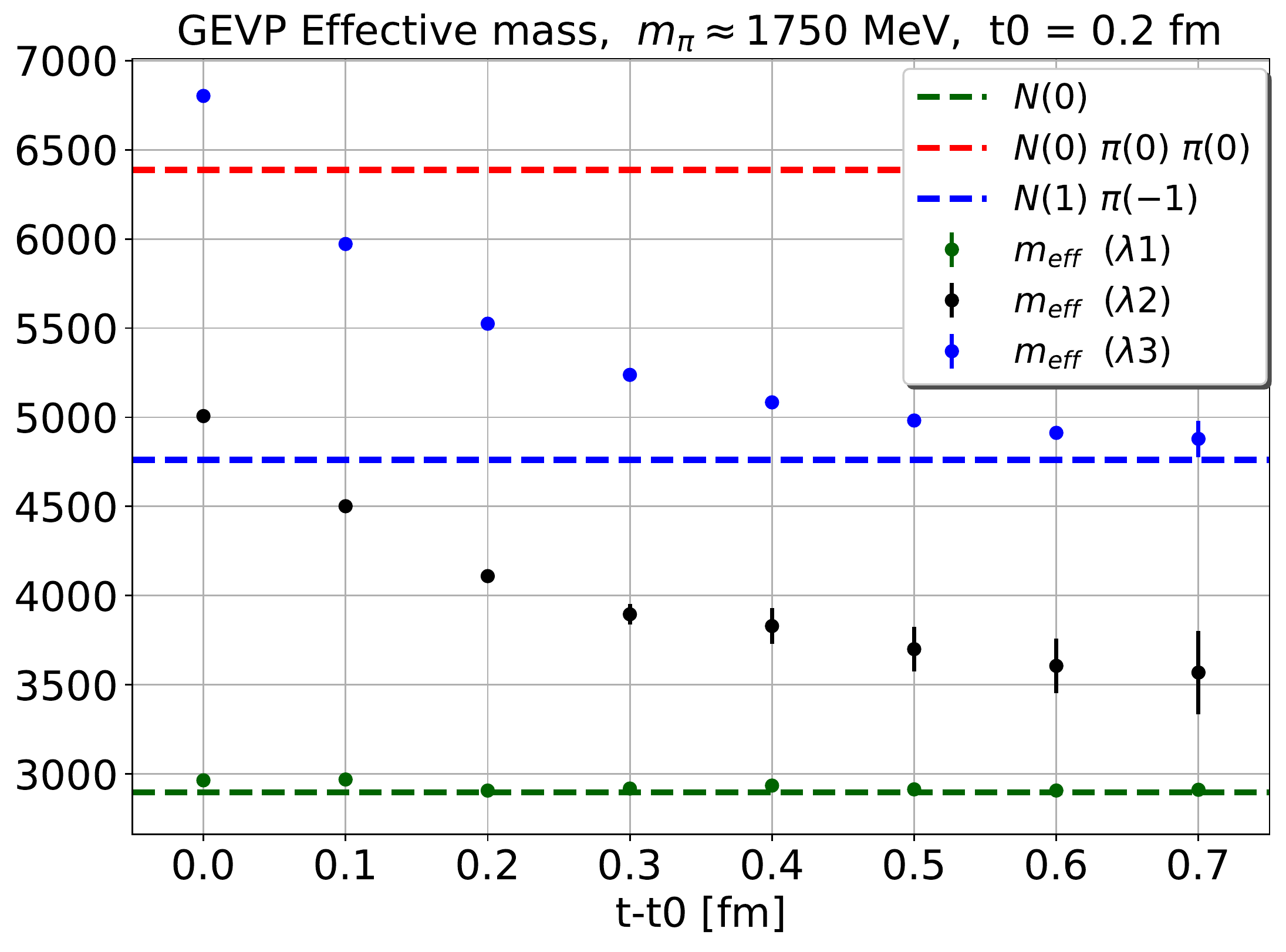} 
\end{minipage}
   \caption{ Effective masses of the eigenvalues extracted with the GEVP analysis using the basis \{${\cal{O}}_N$, $\Phi {\cal{O}}_N$, ${\cal{O}}_{N\pi}$\} for $m_\pi \approx 420$ MeV (left) and $m_\pi \approx 1750$ MeV (right).
   The dashed lines represent the non-interacting energies of the $N$, $N\pi$ and $N\pi\pi$ as in Fig. \ref{figure:meff_nucleon2pt}.
   }
    \label{figure:gevp_energies_3x3}
\end{figure}

If we consider the $p \xrightarrow{\text{${{\cal{J} }}^-$}}p\pi^-$ three-point function 
with a pseudoscalar current ${\cal{J}}^-(z) = \bar{u}(z) \gamma_5 d(z)$ 
and we evaluate it at the source time ($\tau = 0$), 
this corresponds to the two-point function $p\pi^- \to p\pi^-$ at a fixed source-sink separation $t$
(see Fig. \ref{figure:diagrams_p2ppi}).
Therefore, it is possible to construct the two-point function by considering different source-sink separations
for the three-point function. The same holds for the other correlation functions, such that \\
$
C_{2pt}^{p\pi^-\to p\pi^-} (t) ~=~ C_{3pt}^{p \xrightarrow{\text{${{\cal{J} }}^-$}} p\pi^-} (\tau=0; t),
\qquad  \qquad ~
C_{2pt}^{p\pi^-\to n\pi^0} (t) ~=~ C_{3pt}^{p \xrightarrow{\text{${{\cal{J} }}^-$}} n\pi^0} (\tau=0; t),
\\
C_{2pt}^{n\pi^0\to n\pi^0} (t) ~=~ C_{3pt}^{n \xrightarrow{\text{${{\cal{J} }}^0$}} n\pi^0} (\tau=0; t),
\qquad \qquad \quad
C_{2pt}^{n\to p\pi^-} (t) ~=~ C_{3pt}^{n \xrightarrow{\text{${{\cal{J} }}^-$}} p} (\tau=0; t).
$
\\
\hfill \break
After the projection to the isospin component $I_z=-1/2$ (Eq. \eqref{isospin_addition}) and to the irreducible representation $G_1$ (see Eq. \eqref{projection_g1_restframe} for the rest-frame), the correlation functions read:
\begin{align}\label{npi_correlators}
&\braket{{\cal{O}}_{N\pi}^{G_1}(t) ~\bar{\cal{O}}_{N\pi}^{G_1}(0) }=
\frac{2}{3}
\braket{{\cal{O}}^{G_1}_{p\pi^-}(t) ~ \bar{\cal{O}}^{G_1}_{p\pi^-}(0)} 
+
\frac{1}{3}
\braket{{\cal{O}}^{G_1}_{n\pi^0}(t) ~ \bar{\cal{O}}^{G_1}_{n\pi^0}(0)} 
-
\frac{2\sqrt{2}}{\sqrt{3}}
\braket{{\cal{O}}^{G_1}_{p\pi^-}(t) ~ \bar{\cal{O}}^{G_1}_{n\pi^0}(0)} ,
\qquad 
\\
&\braket{{\cal{O}}_{N\pi}^{G_1}(t) ~\bar{\cal{O}}_{n}(0) }~=~
\sqrt{\frac{2}{3}}\braket{{\cal{O}}^{G_1}_{p\pi^-}(t) ~ \bar{\cal{O}}_{n}(0)} ~ 
- \frac{1}{\sqrt{3}}\braket{{\cal{O}}^{G_1}_{n\pi^0}(t) ~ \bar{\cal{O}}_{n}(0)}.
\end{align}
We solved the GEVP for the basis \{${\cal{O}}_N$, $\Phi {\cal{O}}_N$, ${\cal{O}}_{N\pi}$\}
and for the two different values of quark masses.
In Fig. \ref{figure:gevp_energies_3x3} we show the results with $t_0=0.2 ~ \rm{fm}$. 
In the lighter quark mass case, other than the neutron ground state, there is evidence of a state whose energy is similar to the non-interacting $N\pi$ energy in a $P$-wave or a $N\pi\pi$ in a $S$-wave. 
In the heavier quark mass case, there is evidence of a state between the ground state and the non-interacting $N\pi$, which was also present in the GEVP analysis with the basis \{$O_n$, $ \Phi O_{n}$\}.

\section{Conclusions}
In this study,  we used nucleon-pion operators to compute for the first time $N \to N\pi$ three-point functions.
We solved the GEVP using matrices of two-point correlation functions, using the basis \{${\cal{O}}$, $\Phi {\cal{O}}$\}
and \{${\cal{O}}$, $\Phi {\cal{O}}$, ${\cal{O}}_{N\pi}$\} for two quark masses.
With the latter basis and $m_\pi\approx420$ MeV,
we find the nucleon ground state and one state whose energy is close to a non-interacting $N\pi$ in a P-wave or a $N\pi\pi$ in an S-wave.
With $m_\pi\approx1750$ MeV,  in addition we find an intermediate state between the nucleon ground state and the $N\pi$.
We will proceed with the analysis by taking the eigenvectors of the $N$ and $N\pi$ 
calculated from the GEVP for the second basis and using them to extract the matrix elements $\bra{N} {\cal{J}} \ket{N}$ 
and for the first time $\bra{N\pi} {\cal{J}} \ket{N}$ from the three-point functions.

\section{Acknowledgements}
The authors thank S. B\"urger, M. Padmanath, S. Prelovsek, P. Wein and T. Wurm for useful discussion.
This work has received funding from the European Union's Horizon 2020 research and innovation programme under the Marie Skodowska-Curie grant agreement No. 813942 (ITN EuroPLEx).
Simulations were performed on QPACE3 using adapted versions of the software packages \textit{CHROMA} \citep{Edwards_2005}, as well as \textit{GPT} \citep{gpt} for verification.

\bibliographystyle{JHEP}
\bibliography{bibliography}

\providecommand{\href}[2]{#2}\begingroup\raggedright\begin{thebibliography}{10}

\bibitem{Baer_2019_axial}
O.~Bär, \emph{{N}$\pi$-state contamination in lattice calculations of the
  nucleon axial form factors},
  \href{https://doi.org/10.1103/physrevd.99.054506}{\emph{Phys. Rev. D}
  {\bfseries 99} (2019) }.

\bibitem{Baer_2019_pseudoscalar}
O.~Bär, \emph{{N}{$\pi$}-state contamination in lattice calculations of the
  nucleon pseudoscalar form factor},
  \href{https://doi.org/10.1103/physrevd.100.054507}{\emph{Phys. Rev. D}
  {\bfseries 100} (2019) }.

\bibitem{Bali_2020}
G.~Bali et~al., \emph{Nucleon axial structure from lattice {Q}{C}{D}},
  \href{https://doi.org/10.1007/jhep05(2020)126}{\emph{JHEP} {\bfseries 2020}
  (2020) }.

\bibitem{Prelovsek_2017}
S.~Prelovsek et~al., \emph{Lattice operators for scattering of particles with
  spin}, \href{https://doi.org/10.1007/jhep01(2017)129}{\emph{JHEP} {\bfseries
  2017} (2017) }.

\bibitem{Kiratidis:2015vpa}
A.~Kiratidis et~al., \emph{{Lattice baryon spectroscopy with multi-particle
  interpolators}},
  \href{https://doi.org/10.1103/PhysRevD.91.094509}{\emph{Phys. Rev. D}
  {\bfseries 91} (2015) 094509}
  [\href{https://arxiv.org/abs/1501.07667}{{\ttfamily 1501.07667}}].

\bibitem{Leskovec:2018lxb}
L.~Leskovec et~al., \emph{{A lattice {Q}{C}{D} study of pion-nucleon scattering
  in the Roper channel}},
  \href{https://doi.org/10.1007/s00601-018-1419-2}{\emph{Few Body Syst.}
  {\bfseries 59} (2018) 95} [\href{https://arxiv.org/abs/1806.02363}{{\ttfamily
  1806.02363}}].

\bibitem{GUSKEN1990361}
S.~Güsken, \emph{A study of smearing techniques for hadron correlation
  functions},
  \href{https://doi.org/https://doi.org/10.1016/0920-5632(90)90273-W}{\emph{Nucl.
  Phys. B} {\bfseries 17} (1990) 361}.

\bibitem{FALCIONI1985624}
M.~Falcioni et~al., \emph{Again on {S}{U}(3) glueball mass},
  \href{https://doi.org/https://doi.org/10.1016/0550-3213(85)90280-9}{\emph{Nucl.
  Phys. B} {\bfseries 251} (1985) 624}.

\bibitem{Bali:2020lwx}
G.S.~Bali et~al., \emph{{Nonperturbative Renormalization in Lattice {Q}{C}{D}
  with three Flavors of Clover Fermions: Using Periodic and Open Boundary
  Conditions}}, \href{https://doi.org/10.1103/PhysRevD.103.094511}{\emph{Phys.
  Rev. D} {\bfseries 103} (2021) 094511}
  [\href{https://arxiv.org/abs/2012.06284}{{\ttfamily 2012.06284}}].

\bibitem{gupta_axial_2020}
Y.-C.~Jang et~al., \emph{Axial vector form factors from lattice qcd that
  satisfy the pcac relation},
  \href{https://doi.org/10.1103/physrevlett.124.072002}{\emph{Physical Review
  Letters} {\bfseries 124} (2020) }.

\bibitem{Djukanovic_2020}
D.~Djukanovic, \emph{Quark contraction tool — {QCT}},
  \href{https://doi.org/10.1016/j.cpc.2019.106950}{\emph{Comp. Phys. Comm.}
  {\bfseries 247} (2020) 106950}.

\bibitem{Martinelli:1988rr}
G.~Martinelli and C.T.Sachrajda, \emph{{A Lattice Study of Nucleon Structure}},
  \href{https://doi.org/10.1016/0550-3213(89)90035-7}{\emph{Nucl. Phys. B}
  {\bfseries 316} (1989) 355}.

\bibitem{Foster_1999}
M.~Foster and C.~Michael, \emph{Quark mass dependence of hadron masses from
  lattice {Q}{C}{D}},
  \href{https://doi.org/10.1103/physrevd.59.074503}{\emph{Phys. Rev. D}
  {\bfseries 59} (1999) }.

\bibitem{Edwards_2005}
R.G.~Edwards and B.~Joó, \emph{The {C}hroma {S}oftware {S}ystem for {L}attice
  {Q}{C}{D}},
  \href{https://doi.org/10.1016/j.nuclphysbps.2004.11.254}{\emph{Nucl. Phys. B}
  {\bfseries 140} (2005) 832–834}.

\bibitem{gpt}
C.~Lehner, ``{GRID} {P}ython {T}oolkit ({GPT}).''
  \url{http://github.com/lehner/gpt.git}, 2020.

\end{thebibliography}\endgroup

\end{document}